\documentclass[a4,12pt]{article}

\usepackage{bm}
\usepackage[dvipdfmx]{graphicx}
\usepackage{amsmath}
\usepackage{amsfonts}

\newcommand{\ol}{\overline}
\newcommand{\wt}{\widetilde}

\newcommand{\ZZ}{\mathbb{Z}}
\newcommand{\RR}{\mathbb{R}}

\def\Re{\mathop{\rm Re}\nolimits}
\def\Im{\mathop{\rm Im}\nolimits}

\begin{document}
\begin{titlepage}
\title{
\vspace{-2cm}
\begin{flushright}
\normalsize{
TIT/HEP-655\\
KCL-MTH-16-03\\
June 2016}
\end{flushright}
       \vspace{1cm}
       Supersymmetry Enhancement and Junctions in S-folds
       \vspace{1cm}}
\author{
Yosuke Imamura\thanks{E-mail: \tt imamura@phys.titech.ac.jp}$^{~1}$,
Hirotaka Kato\thanks{E-mail: \tt h.kato@th.phys.titech.ac.jp}$^{~1}$,
and Daisuke Yokoyama\thanks{E-mail: \tt daisuke.yokoyama@kcl.ac.uk}$^{~2}$
\\[30pt]
{\it $^1$ Department of Physics, Tokyo Institute of Technology,}\\
{\it Tokyo 152-8551, Japan}\\
{\it $^2$ Department of Mathematics, King's College London,}\\
{\it The Strand, London, WC2R 2LS, U.K.}
}
\date{}

\thispagestyle{empty}

\vspace{0cm}

\maketitle

\begin{abstract}
We study supersymmetry enhancement from ${\cal N}=3$ to ${\cal N}=4$
proposed by Aharony and Tachikawa by using string junctions in S-folds.
The central charges carried by junctions play a central role in our analysis.
We consider planer junctions in a specific plane.
Before the S-folding they carry two complex central charges,
which we denote by $Z$ and $\ol Z$.
The S-fold projection eliminates $\ol Z$ as well as one of the four supercharges,
and when the supersymmetry is enhanced $\ol Z$ should be reproduced
by some non-perturbative mechanism.
For the models of $\ZZ_3$ and $\ZZ_4$ S-folds
which are expected to give $SU(3)$ and $SO(5)$ ${\cal N}=4$ theories
we compare the junction spectra with those in perturbative brane realization of the same theories.
We establish one-to-one correspondence
so that $Z$ coincides.
By using the correspondence we also give an expression for the
enhanced central charge $\ol Z$.
\end{abstract}
\end{titlepage}

\section{Introduction}
A superconformal algebra in four dimensions
is specified by the number of supercharges, ${\cal N}\leq4$.
Theories with ${\cal N}=1,2,4$ have been studied extensively
in the last few decades,
and played essential roles in the development
of field theories.
${\cal N}=3$ theories, however,
had not attracted much attention until quite recently.
A reason for this is that
the only ${\cal N}=3$ multiplet of free fields
is the vector multiplet, which has the same
contents as the ${\cal N}=4$
vector multiplet, and
if we try to make an ${\cal N}=3$ theory
in a perturbative way
we always end up with an ${\cal N}=4$ theory.
This means if there exist genuine ${\cal N}=3$ theories
they should necessarily be strongly coupled.
This is consistent to the fact that the
absence of marginal deformations in
${\cal N}=3$ theories \cite{Aharony:2015oyb,Cordova:2016xhm}.

An explicit construction of ${\cal N}=3$ theories was recently proposed
by Garc\'ia-Etxebarria and Regalado in \cite{Garcia-Etxebarria:2015wns}%
\footnote{See also \cite{Ferrara:1998zt} for an attempt of studying ${\cal N}=3$ theories
by using AdS/CFT correspondence.}.
They realize ${\cal N}=3$ theories on a stack of D3-branes
by performing a $\ZZ_k$ orbifolding that project out one of four supercharges
on the D3-branes.
Such an orbifold can be defined
by combining
$U(1)_R$ symmetry of
type IIB supergravity \cite{Schwarz:1983qr,Schwarz:1983wa}
and a rotation of the transverse space $\RR^6$.
The $U(1)_R$ is broken by the quantization of $(p,q)$ string charges
down to a discrete subgroup depending on the value of the complex field $\tau=\chi+ie^{-\phi}$.
For generic $\tau$ the subgroup is $\ZZ_2$ and then the orbifolded theory is still ${\cal N}=4$,
while for some special values $\ZZ_k$ with $k=3,4,6$ are unbroken and then we can
define genuine ${\cal N}=3$ theories.
The resulting orbifolds, which we call S-folds following \cite{Aharony:2016kai},
are labelled by two integers $k$ and $\ell$;
$k$ is the order of the orbifold group
and $\ell$ is a divisor of $k$ related to the discrete torsion.
Therefore, the theory on the D3-branes are labeled by $k$, $\ell$, and $N$,
where $N$ is the number of mobile D3-branes.

Aharony and Tachikawa
\cite{Aharony:2016kai} 
listed the dimensions of the Coulomb branch operators
in these theories\footnote{In \cite{Aharony:2016kai} it is pointed out that the theories on the S-folds are obtained by gauging certain discrete symmetries of parent theories.
A parent theory includes operators charged under the discrete symmetry, and in the corresponding gauged theory they are projected out and do not appear in the spectrum of gauge invariant operators.
The current of enhanced supersymmetry is such an operator, and the theories we study in this paper are not the gauged theories but parent ones.
Our analysis in the following section is not affected by this difference.
}.
In particular, they contain an operator with dimension $N\ell$,
which becomes $1$ or $2$ for
some of the theories.
Aharony and Evtikhiev showed in \cite{Aharony:2015oyb}
that
the dimensions of Coulomb branch operators
in a genuine ${\cal N}=3$ theory
must be equal to or greater than $3$.
This was confirmed for rank-$1$ ${\cal N}=3$ theories
in \cite{Nishinaka:2016hbw,Argyres:2016xua}.
This fact implies that
the ${\cal N}=3$ theories with $N\ell=1$ or $2$
are in fact ${\cal N}=4$ theories \cite{Aharony:2016kai},
which can be specified by giving the gauge group $G$.
In \cite{Aharony:2016kai} 
 some BPS states were identified with $1/2$ BPS W-bosons,
and the gauge groups were determined (Table \ref{n4list}).

\begin{table}[htb]
\caption{Theories in which ${\cal N}=3$ is enhanced to ${\cal N}=4$ are shown.}
\label{n4list}
\vspace{10pt}
\begin{center}
\begin{tabular}{cc}
\hline
\hline
$(k,\ell,N)$ & $G$ \\
\hline
$(k,1,1)$ & $U(1)$ \\
\hline
$(3,1,2)$ & $SU(3)$ \\
$(4,1,2)$ & $SO(5)$ \\
$(6,1,2)$ & $G_2$ \\
\hline
\end{tabular}
\end{center}
\end{table}

The purpose of this paper is to analyze the spectra
of these theories.
Because each theory is defined by a system of D3-branes in an S-fold,
we can realize dyonic particles by string junctions \cite{Schwarz:1996bh,Dasgupta:1997pu,Sen:1997xi,Rey:1997sp,Bergman:1997yw} connecting
D3-branes and the S-plane.

This analysis is highly
non-trivial.
Let us consider $(k,\ell,N)=(k,1,1)$ case.
As is pointed out in \cite{Aharony:2016kai},
this should give the ${\cal N}=4$ Maxwell theory.
We can consider
an open $(p,q)$ string connecting
a D3-brane and one of its mirror branes.
Although it seems to give massive BPS states
with non-vanishing dyonic charge
in a generic point of the Coulomb branch
there are no such massive BPS states
in the ${\cal N}=4$ Maxwell theory.
This implies that the string should be decoupled in the
field theory limit.
One may think that
the central charge of a string is
proportional to its length,
and we should have light massive BPS states when
the D3-brane is sufficiently close to the S-plane.
However, the central charge gives only a
lower bound of the mass,
and does not guarantee the existence of the light states.
We consider planar junctions in a specific plane.
Before the $\ZZ_k$ S-folding
we can define two complex central charges $Z$ and $\ol Z$ carried
by junctions.
One of them, say $\ol Z$, is projected out in the S-fold.
In the system with the supersymmetry enhancement,
the central charge $\ol Z$ revives non-perturbatively,
and it may give a bound greater than the perturbative one.
In this case the naive expectation about
the presence of light BPS states
would fail.

Unfortunately, in the present paper,
we cannot directly determine the
non-perturbative bound.
Instead, we determine it
by using another realization of the theory in which the ${\cal N}=4$ symmetry is manifest.
Some of the gauge groups in Table \ref{n4list}
are classical Lie groups,
and such ${\cal N}=4$ theories have a simpler brane
realization,
in which ${\cal N}=4$ supersymmetry
and the associated central charges are realized perturbatively.
We first compare the central charges $Z$ that is realized perturbatively
both in the S-fold and in the perturbative set-up.
With this information
we establish
relations between Coulomb moduli
and dyonic charges in the S-fold
and those in the other set up.
Then, by assuming the equivalence of the two theories,
we guess an expression of
the non-perturbative central charge $\ol Z$
in terms of the Coulomb moduli and the dyonic charges
of the particle.

This paper is organized as follows.

In the next section we study general aspects of
S-folds and junctions in them.
Constraints imposed on the charges of strings ending on S-planes
and the perturbative BPS bound in S-folds are given.
In Section \ref{matching.sec}
we give one-to-one correspondence between
the charge lattice of junctions
in the S-fold realization and
that in a perturbative realization
of ${\cal N}=4$
$SU(3)$ and $SO(5)$ theories.
The structure of a lattice is related non-trivially
to the properties of the S-plane or O-plane which exists in the set-up,
and the successful matching
seems to strongly support the conjecture
of the supersymmetry enhancement.
Furthermore, we propose a formula to determine central charges without
referring to the simpler realization.
It gives $Z$ and $\ol Z$
even in the $G_2$ case,
which does not have perturbative string realization.

In Section \ref{wall.sec} we carry out a preliminary analysis of
walls of marginal stability \cite{Bergman:1997yw}.
We focus on strings in the $\ZZ_3$ and $\ZZ_4$ S-folds
that connect a D3-brane and one of its mirror,
and determine walls in Coulomb branch moduli space
based on the assumption of the supersymmetry enhancement.
The results are different from what are expected
by perturbative analysis on the S-folds.

Section \ref{discussions.sec} is devoted to discussions.

\section{Junctions and BPS conditions}
\subsection{Junctions in S-folds}
In this subsection we will consider junctions in S-folds.

Let us first consider junctions in a system
of $n$ parallel D3-branes
in the flat background without any S-folding.
A junction $\bm{j}$ is specified by charges
$(p_a,q_a)$ $(a=1,\ldots,n)$
of strings ending on the D3-branes.
For strings ending on mobile D3-branes we always
define string charges as incomming charges.
When we discuss S-folds later it is more convenient to use the
complex charges $Q_a=p_a+\tau q_a$.
The charge conservation requires
$Q_a$ to satisfy
\begin{align}
\sum_{a=1}^nQ_a=0.
\label{totalq}
\end{align}
When the positions of D3-branes are generic
the low-energy effective theory of the parallel
D3-branes is the ${\cal N}=4$ $U(1)^n$ supersymmetric
gauge theory.
Let $U(1)_a$ be the gauge group corresponding to the $a$-th D3-brane.
In the low-energy effective theory
a junction is regarded as a particle with
the $U(1)_a$ dyonic charge $Q_a$.

Let us consider junctions in $\ZZ_k$ S-fold.
For $k\geq 3$ we set $\tau=\gamma\equiv e^{2\pi i/k}$,
while for $k=2$ we can take an arbitrary $\tau$.
Let $\Gamma_k$ be the lattice on the complex plane spanned by $1$ and $\tau$.

We introduce $N$ D3-branes on the S-fold.
As in the case of orientifolds it may be possible to introduce
D3-branes that are mirror to themselves.
We introduce them later as a non-trivial discrete torsion.
Here we assume the absence of such trapped D3-branes at the
fixed point.
Then in the covering space we have $kN$ D3-branes.
We label them by two indices $a=1,\ldots,N$ and $i=0,\ldots,k-1$,
where $a$ labels $N$ independent branes and $i$ labels $k$ branes identified by $\ZZ_k$ action.
Their positions $z_{i,a}$
are related by
\begin{align}
z_{i,a}=\gamma^i z_a,
\end{align}
where $z_a\equiv z_{0,a}$.
As in the flat background a junction
is specified by the string charges $Q_{i,a}\in\Gamma_k$
on the D3-branes.
We define $\ZZ_k$ action so that it rotates
the coordinate $z$ and the complex charge $Q$ in the opposite
direction by the same angle as given in (\ref{zqrotation}).
Due to the $\ZZ_k$ identification a string with charge $Q_{i,a}$ attached on the brane at $z_{i,a}$
is equivalent to a string with charge $\gamma^iQ_{i,a}$ attached on the brane at $z_a$.
When we read off the $U(1)^N$ dyonic charges $Q_a$
we should take account of this equivalence.
We collect the endpoints on $z_{i,a}$ with $i=0,\ldots,k-1$
to the brane with $i=0$ by $\ZZ_k$ transformations
and obtain
\begin{align}
Q_a=\sum_{i=0}^{k-1}\gamma^iQ_{i,a}.
\label{qadef}
\end{align}
These charges are again elements of $\Gamma_k$.
We have assumed trivial discrete torsion,
and there are no D3-branes at the origin of the covering space.
In this case a junction is attached on only mobile D3-branes and
due to the charge conservation $Q_{i,a}$ must satisfy
\begin{align}
\sum_{i,a}Q_{i,a}=0.
\label{sumqiazero}
\end{align}
Due to this constraint
$Q_a$ defined by (\ref{qadef}) may not be all independent.
The constraint is obtained as follows.

Let $F_k$ be a $\ZZ_k$ invariant homomorphism from $\Gamma_k$ onto
some discrete group $K$.
The $\ZZ_k$ invariance means $F_k(\gamma Q)=F_k(Q)$
for an arbitrary $Q\in\Gamma_k$%
\footnote{The group $K$ is easily obtained as follows.
Let us consider $k=2$ case as an example.
Then $\Gamma_2$ is the lattice generated by $1$ and $\tau$.
Let $a$ and $b$ be two elements of $K$ corresponding
to them: $a=F_2(1)$ and $b=F_2(\tau)$.
By the linearity $F_2(p+q\tau)=pa+qb$ ($p,q\in\ZZ$),
and the $\ZZ_2$ invariance requires $2a=2b=0$.
This means $K$ contains at most four elements
$0$, $a$, $b$, and $a+b$.
If these are all different we obtain the maximal $K$
($=\ZZ_2+\ZZ_2$).
If one of $a$, $b$, and $a+b$ is zero $K=\ZZ_2$,
and if $a=b=0$ then $K=0$.}.
By using (\ref{qadef}),
(\ref{sumqiazero}), and the $\ZZ_k$ invariance of $F_k$
we can easily show
\begin{align}
F_k(Q_0)=0,
\label{fq0}
\end{align}
where $Q_0$ is the total charge
\begin{align}
Q_0=\sum_{a=1}^NQ_a.
\label{q0}
\end{align}
If $K$ is not trivial (\ref{fq0}) gives a non-trivial constraint
imposed on the set of the charges $Q_a$.
On the S-fold, $Q_0$
can be regarded as the charge of a string attached on
the S-plane.
(For S-planes and O-planes we define the charge of a string
ending on them as the outgoing charge.)
Therefore,
(\ref{fq0}) gives the constraint on
the string charge that can be attached on the S-plane.

The constraint (\ref{fq0}) is a necessary condition.
There may be some different choices of $K$, and
for some of them (\ref{fq0}) may not be a sufficient one.
However, we can easily show that
if we choose maximal $K$
(\ref{fq0}) gives the sufficient condition
for the existence of junctions.
The maximal $K$ are in fact isomorphic to the discrete torsion group
\begin{align}
\Gamma^{\rm tor}=H^3(\bm{S}^5/\ZZ_k,\wt{\ZZ+\ZZ})
\end{align}
for the $\ZZ_k$ S-fold \cite{Witten:1998xy,Aharony:2016kai,Imamura:2016abe}
(Table \ref{torsion}).
\begin{table}[htb]
\caption{The discrete torsion groups for $\ZZ_k$ S-folds are shown.}\label{torsion}
\begin{center}
\begin{tabular}{ccccc}
\hline
\hline
$k$ & $2$ & $3$ & $4$ & $6$ \\
\hline
$\Gamma^{\rm tor}$ & $\ZZ_2+\ZZ_2$ & $\ZZ_3$ & $\ZZ_2$ & $0$ \\
\hline
\end{tabular}
\end{center}
\end{table}

Up to here we consider S-folds with the trivial discrete torsion.
Introduction of a non-vanishing discrete torsion changes the condition
given above.
To determine the modified condition
it is convenient to realize such an S-plane by using
wrapped fivebranes.
(Similar realization of O-planes with non-trivial discrete torsions was given in
\cite{Witten:1998xy,Hyakutake:2000mr}.)
By the Poincare duality the discrete torsion
can be also regarded as the two-cycle homology.
Let us consider an S-fold with a torsion $t\in\Gamma^{\rm tor}$.
We can realize such an S-fold from that with the trivial torsion
by wrapping a fivebrane around the two-cycle
specified by $t$.
(Note that the coefficient group $\wt{\ZZ+\ZZ}$ is the sheaf of the
$(p,q)$ charges of fivebranes and an element of $\Gamma^{\rm tor}$ specifies
not only the cycle wrapped by the fivebrane but also the fivebrane charges.)
The existence of the wrapped $(p,q)$ fivebrane
allows $(p,q)$ strings to end on it.
Namely, string with charge $Q_0$ can be attached on the S-plane
if
\begin{align}
F(Q_0)\in \ZZ t.
\label{attachable}
\end{align}
The right hand side is not just $t$ but $\ZZ t$ because
we can attach an arbitrary number of strings.

\subsection{BPS bounds}
\label{sec:bps-conditions}
Let us consider supersymmetry and BPS bounds in
${\cal N}=4$ theories.
We denote the four supercharges with the positive chirality
by $Q_\alpha^I$
and their Hermitian conjugate by $\ol Q_{\dot\alpha I}$.
The indices $\alpha=1,2$ and $\dot\alpha=\dot 1,\dot 2$ are left-handed and right-handed spinor indices
for $SO(1,3)_{0123}$, respectively.
The upper and lower indices $I=1,2,3,4$ label the fundamental and the anti-fundamental
representations of $SU(4)_R\sim SO(6)_{456789}$.
Their anti-commutation relations are
\begin{align}
\{Q^I_\alpha,\ol Q_{\dot\beta J}\}
=\sigma^\mu_{\alpha\dot\beta}\delta^I_JP_\mu,\quad
\{Q_\alpha^I,Q_\beta^J\}=\epsilon_{\alpha\beta}\rho_m^{IJ}Z^{m+3},
\label{n4algebra}
\end{align}
where
$Z^{m+3}$ ($m=1,\ldots,6$) are central charges belonging to
the vector representation of $SU(4)_R$, and
$\rho_m^{IJ}$ is an $SU(4)_R$ invariant tensor
with two anti-symmetric fundamental indices and one vector index.
The central charges $Z^{m+3}$ form a complex vector in the $\RR^6$
transverse to the D3-branes,
and its real and imaginary part 
can be interpreted
as the extension of fundamental strings
and that of D-strings, respectively,
in the brane realization of the ${\cal N}=4$ SYM.

The BPS bound obtained from (\ref{n4algebra}) is \cite{Fraser:1997nd}
\begin{align}
m^2\geq|\Re\vec Z|^2+|\Im\vec Z|^2
+|\Re\vec Z||\Im\vec Z|\sin\alpha,
\end{align}
where $\alpha$ is the angle between two vectors
$\Re\vec Z$ and $\Im\vec Z$ in $\RR^6$.
For a junction to be BPS and to saturate this bound
it must be planar \cite{Fraser:1997nd,Lee:1998nv}.
We consider planer junctions in $89$ plane, and
it is convenient to decompose the
$SO(6)_{456789}$ fundamental representation
into $SO(4)_{4567}\times SO(2)_{89}$
representations as
\begin{align}
\bm{4}=(\bm{2},\bm{1})_{+\frac{1}{2}}+(\bm{1},\bm{2})_{-\frac{1}{2}}.
\end{align}
We denote the supercharges belonging to
$(\bm{2},\bm{1})_{+1/2}$ and $(\bm{1},\bm{2})_{-1/2}$ by
$Q_\alpha^a$ and $Q_\alpha^{\dot a}$, respectively,
and their Hermitian conjugate by
$\ol Q_{\dot\alpha a}$ and $\ol Q_{\dot\alpha\dot a}$, respectively.
If only two components $Z^8$ and $Z^9$ out of the six are non-vanishing,
$Q_\alpha^a$
and $Q_\alpha^{\dot a}$
anticommute
to each other.
Then the algebra (\ref{n4algebra}) splits into two copies of ${\cal N}=2$ algebra.
\begin{align}
\{Q_\alpha^a,\ol Q_{\dot\beta b}\}=\sigma^\mu_{\alpha\dot\beta}\delta^a_b P_\mu,\quad
\{Q_\alpha^a,Q_\beta^b\}
=\epsilon_{\alpha\beta}\epsilon^{ab}Z,\nonumber\\
\{Q_\alpha^{\dot a},\ol Q_{\dot\beta\dot b}\}=\sigma^\mu_{\alpha\dot\beta}\delta^{\dot a}_{\dot b} P_\mu,\quad
\{Q_\alpha^{\dot a},Q_\beta^{\dot b}\}
=\epsilon_{\alpha\beta}\epsilon^{\dot a\dot b}\ol Z,
\label{qqzz}
\end{align}
where $Z=Z^8+iZ^9$ and $\ol Z=Z^8-iZ^9$.
Note that $Z^8$ and $Z^9$ are complex,
and $Z$ and $\ol Z$ are not conjugate to each other.
$Z$ and $\ol Z$ carry $U(1)_R\times SO(2)_{89}$ charges
$(+1,+1)$ and $(+1,-1)$, respectively.
For an open $(p,q)$ string with the complex charge
$Q=p+\tau q$ and the extension $\Delta z$ on the $89$ plane
these central charges are given by
\begin{align}
Z=Q\Delta z,\quad
\ol Z=Q\Delta z^*.
\end{align}
(Note that $Q$ and $\Delta z$ carry $SO(2)_R\times SO(2)_{89}$ charges $(+1,0)$ and $(0,+1)$,
respectively.)
Each ${\cal N}=2$ algebra
in (\ref{qqzz}) gives the bound independently:
\begin{align}
m\geq|Z|,\quad
m\geq|\ol Z|.
\label{twobounds}
\end{align}

For a string junction consisting of more than one open string
the central charges of the junction are given by
\begin{align}
Z=\sum_iQ_iz_i,\quad
\ol Z=\sum_iQ_iz_i^*,
\label{zisqz}
\end{align}
where $z_i$ are the positions of D3-branes and $Q_i$ are complex charges of strings ending on them.
If a junction contains more than two strings with different charges only one of the bounds
in (\ref{twobounds}) can be saturated.
To saturate the bound $m\geq|Z|$,
all the constituent open strings
should satisfy $m=|Z|$.
This means that $\arg Z$ for the strings must be the same.
Namely, the angles $\arg Q+\arg\Delta z$ are the same for all strings.
We call such a junction ``a holomorphic junction.''
On the other hand,
for a junction saturating the bound $m\geq|\ol Z|$,
the angles $\arg Q-\arg\Delta z$ are the same for all the the constituent strings.
We call such a junction ``an anti-holomorphic junction.''%
\footnote{These two types of junctions are called class A and class B
in \cite{Sen:2012hv}.}
Two bounds can be saturated at the same time only for
a set of parallel strings with the same charge.
Every state belongs to one of the following four types:
\begin{itemize}
\item $m=|Z|=|\ol Z|$: $1/2$ BPS states
\item $m=|Z|>|\ol Z|$: holomorphic $1/4$ BPS states
\item $m=|\ol Z|>|Z|$: anti-holomorphic $1/4$ BPS states
\item $m>|Z|$ and $m>|\ol Z|$: non-BPS states.
\end{itemize}


Next, let us consider how these BPS bounds in ${\cal N}=4$ theory
are modified when we perform an S-fold projection.
An S-fold is defined by the projection which leaves states invariant under the $\ZZ_k$ action
generated by
\begin{align}
g=\gamma^{J_{45}-J_{67}+J_{89}-R},
\end{align}
where $J_{ij}$ and $R$ are generators of $SU(4)_R$ and $U(1)_R$, respectively.
$g$ acts on the complex coordinate of the $89$-plane $z$ and
the complex string charge $Q$ as
\begin{align}
z\stackrel{g}{\longrightarrow}z'=\gamma z,\quad
Q\stackrel{g}{\longrightarrow}Q'=\gamma^{-1}Q.
\label{zqrotation}
\end{align}
The supercharges $Q^a$ and $Q^{\dot a}$ and
the central charges $Z$ and $\ol Z$ are transformed as
\begin{align}
(Q^1,Q^2,Q^{\dot 1},Q^{\dot 2})
\rightarrow (Q^1,Q^2,Q^{\dot 1},\gamma^{-2}Q^{\dot 2}),\quad
Z\rightarrow Z,\quad
\ol Z\rightarrow \gamma^{-2}\ol Z.
\end{align}
If $k\geq 3$ $Q^{\dot 2}$ and $\ol Z$ are projected out
and we obtain ${\cal N}=3$ theory.
This theory is generically genuine ${\cal N}=3$ theory.
However, there are cases that
an additional supersymmetry is generated non-perturbatively,
and what we are interested in are such cases.
Because the algebra of this emergent ${\cal N}=4$ symmetry is isomorphic
to the usual one, we have the same BPS bounds $m\geq |Z|$ and $m\geq|\ol Z|$.
However, $\ol Z$ is not given by (\ref{zisqz}).

Before ending this section, we give a comment on the existence of
BPS saturating states.
For the ${\cal N}=4$ $SU(N)$ SYM the BPS spectrum
of dyons and the corresponding junctions have been studied in detail.
A particle corresponding to a $(p,q)$ string 
is $1/2$ BPS and its ground states form a short multiplet.
A particle realized as a $3$-pronged junction
has $12$ fermionic zero-modes \cite{Bergman:1998gs,Lee:1998nv,Bak:1999hp}
and its ground states form a multiplet of middle size.
More general junctions have also been studied
and, for example, a prescription to calculate
their contribution to some BPS indices
is known \cite{Sen:2012hv},
and generically we have BPS saturating states.

However, in more general
situation with S-plane or O-plane,
this is not the case.
Even if we can draw a junction whose constituent strings
satisfy the condition we mentioned above,
it does not mean the existence of BPS quantum states.
For example let us consider a fundamental
string connecting a D3-brane
and its mirror in an orientifold background.
Although it seems $1/2$ BPS, it is indeed the case only when the
RR-charge of the O-plane is positive.
If the RR-charge is negative
the BPS saturating states
are projected out by the orientifold projection,
and the mass of the string is on the order of string scale
even if the D-branes coincide with the O-plane.

\section{Matching of charges and moduli}\label{matching.sec}
Let us consider junctions in the $\ZZ_k$ S-fold with
trivial discrete torsion.
If we put two D3-branes
the supersymmetry is expected to be enhanced from ${\cal N}=3$ to ${\cal N}=4$
\cite{Aharony:2016kai}.
In this section we determine the both central charges $Z$ and $\ol Z$ for junctions
in the S-folds
by using relations to perturbative realization of the same ${\cal N}=4$ theories.

Following \cite{Aharony:2016kai} let us begin with
the identification of W-bosons in the $\ZZ_k$ S-fold.
We consider an open string connecting two D3-branes.
We denote a string with complex charge $Q$ that goes from a D3-brane at $z_{j,2}$ to another D3-brane at $z_{i,1}$ by
\begin{align}
z_{i,1}\stackrel{Q}{\longleftarrow} z_{j,2},\quad
Q\in\Gamma_k.
\label{wbv}
\end{align}
Let $Q_{i,a}^{(S)}$ be the charges of strings attached on D3-branes in the covering space.
(We put the superscript ``$(S)$'' for distinction from the quantities in the dual set-up
we will introduce later.)
This is the junction with $Q^{(S)}_{i,1}=Q$, $Q^{(S)}_{j,2}=-Q$, and the other
$Q^{(S)}_{i,a}$ vanishing.
(\ref{qadef}) gives
\begin{align}
(Q^{(S)}_1,Q^{(S)}_2)=(\gamma^iQ,-\gamma^jQ).
\label{qs1qs2}
\end{align}
Following \cite{Aharony:2016kai} we impose the following electric condition:
\begin{align}
Q^{(S)}_1=-Q^{(S)*}_2.
\label{electric}
\end{align}
Then the string charge $Q$ must satisfy
\begin{align}
Q=\pm|Q|\gamma^{-\frac{i+j}{2}},
\end{align}
and
(\ref{qs1qs2}) becomes
\begin{align}
(Q^{(S)}_1,Q^{(S)}_2)=\pm|Q|(\gamma^{\frac{i-j}{2}},-\gamma^{\frac{j-i}{2}}).
\label{wbosoncharges}
\end{align}
Different $i-j$ gives different charges, and it is shown in \cite{Aharony:2016kai}
that these charges form the weight vectors in the adjoint representation
of a rank-$2$ Lie group depending on $k$
(Table \ref{n4list}).

In the following subsections, we give detailed analysis of
the three models with supersymmetry enhancement.

\subsection{$\ZZ_3$ S-fold}
\begin{figure}[htb]
\begin{center}
\includegraphics{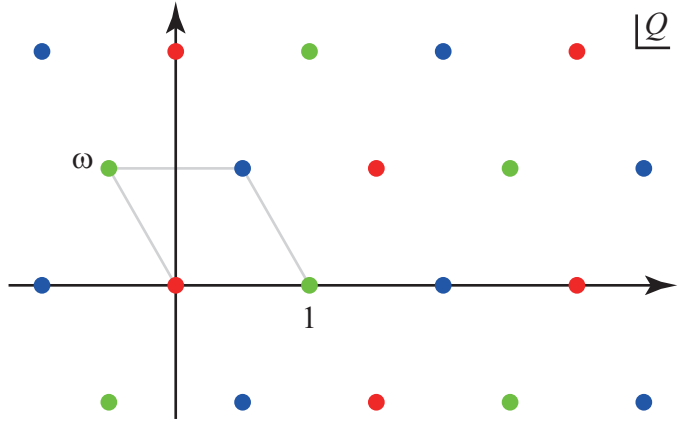}
\end{center}
\caption{The $\ZZ_3$ grading of the lattice $\Gamma_3$ is shown.
Red, green, and blue dots represent charges with $F(Q)=0$, $1$ and $2$, respectively.}\label{z3grading.eps}
\end{figure}

Let us first consider $\ZZ_3$ S-fold.
In this case the modulous on the S-fold is $\tau^{(S)}=\omega=e^{2\pi i/3}$,
and the complex string charges form the triangular lattice $\Gamma_3$ in Figure \ref{z3grading.eps}.
(For $k=3$ we use $\omega$ instead of $\gamma$.)
The charges $Q_1^{(S)}$ and $Q_2^{(S)}$ are elements of $\Gamma_3$,
and constrained by (\ref{fq0}),
\begin{align}
F_3(Q_1^{(S)}+Q_2^{(S)})=0,
\label{z3constraint}
\end{align}
where $F_3$ is the map from $\Gamma_3$ onto $\ZZ_3$.
This defines the $\ZZ_3$ grading in the lattice shown in Figure \ref{z3grading.eps}.
A string can end on the S-plane only when its charge
is a red dot in Figure \ref{z3grading.eps}.

Different choices of $j$ and $i$ in (\ref{wbosoncharges})
with the same $j-i$ give
equivalent strings mapped to one another by $\ZZ_k$, and
only $j-i$ is significant.
Let $\bm{a}$, $\bm{b}$, and $\bm{c}$ denote the
strings corresponding to $j-i=0$, $1$, and $2$.
In the $\ZZ_3$-frame in which the string charge becomes $Q=1$,
these strings are
\begin{align}
\bm{a}:z_1\stackrel{1}{\longleftarrow}z_2,\quad
\bm{b}:\omega z_1\stackrel{1}{\longleftarrow}\omega^2 z_2,\quad
\bm{c}:\omega^2z_1\stackrel{1}{\longleftarrow}\omega z_2.
\label{ABC}
\end{align}
The dyonic charges and central charges of these strings are shown in Table \ref{z3charges}.
\begin{table}[htb]
\caption{The dyonic charges and the central charge of junctions in the S-fold and flat background
are shown for three strings corresponding to W-bosons.}\label{z3charges}
\begin{center}
\begin{tabular}{ccccc}
\hline
\hline
         & $(Q_1^{(S)},Q_2^{(S)})$ & $Z^{(S)}$ & $(Q_1^{(F)},Q_2^{(F)})$ & $Z^{(F)}$ \\
\hline
$\bm{a}$ & $(1,-1)$ & $z_1-z_2$ & $(1,-1)$ & $y_1-y_2$ \\
$\bm{b}$ & $(\omega,-\omega^2)$ & $\omega z_1-\omega^2 z_2$ & $(-1,0)$ & $-y_1$ \\
$\bm{c}$ & $(\omega^2,-\omega)$ & $\omega^2 z_1-\omega z_2$ & $(0,1)$ & $y_2$ \\
\hline
\end{tabular}
\end{center}
\end{table}
As is confirmed in \cite{Aharony:2016kai}
these are the same as those of W-bosons of $SU(3)$ SYM.

To see this more clearly,
let us use another realization
of the ${\cal N}=4$ $SU(3)$ theory
with parallel D3-branes
in the flat background without any S-folding.
We use $y$ for the coordinate of D3-branes.
The center of mass motion is decoupled and we take the coordinate so that
one of the D3-branes is located at $y=0$,
and denote the positions of the other two branes by $y_1$ and $y_2$.
We denote the charges of the strings
ending at $y_1$ and $y_2$
by $Q_1^{(F)}$ and $Q_2^{(F)}$, respectively.
(The superscripts ``$(F)$'' indicate the flat background.)
By using these charges we can determine the central charge $Z^{(F)}$ by the
general formula (\ref{zisqz}).

Let us assume that the strings in (\ref{ABC})
correspond to fundamental strings stretched between
two D3-branes on the $y$-plane.
The sum of the charges of the three strings in (\ref{ABC})
vanishes, and
we choose the corresponding strings
so that they also carry charges with vanishing sum:
\begin{align}
\bm{a}: y_1\stackrel{1}{\longleftarrow}y_2,\quad
\bm{b}: 0\stackrel{1}{\longleftarrow}y_1,\quad
\bm{c}: y_2\stackrel{1}{\longleftarrow}0.
\label{ABC2}
\end{align}
The central charge $Z^{(F)}$ and the dyonic charges $(Q_1^{(F)},Q_2^{(F)})$ for these are
shown in Table \ref{z3charges}.
Because
all strings in (\ref{ABC2}) are fundamental ones,
the central charge does not depend on the modulous $\tau^{(F)}$ on the
flat background, which is not fixed yet.

We require $Z^{(S)}=Z^{(F)}$ for the W-boson states in Table \ref{z3charges}.
This means
\begin{align}
Z=Q_1^{(F)}y_1+Q_2^{(F)}y_2=Q_1^{(S)}z_1+Q_2^{(S)}z_2.
\end{align}
By requiring this equality for the W-bosons
we obtain the relations among $z_a$ and $y_a$:
\begin{align}
z_1=\frac{\omega y_1+\omega^2 y_2}{\omega-\omega^2},\quad
z_2=\frac{\omega^2 y_1+\omega y_2}{\omega-\omega^2}.
\label{z1z2byy1y2}
\end{align}

$Z^{(S)}$ and $Z^{(F)}$ are holomorphic functions in the complex charges.
This means that $Q_a^{(S)}$ and $Q_a^{(F)}$ should be related by holomorphic linear relations.
The comparison of the complex charges of W-boson states in Table \ref{z3charges}
determines the following relations.
\begin{align}
Q_1^{(F)}=\frac{\omega Q_1^{(S)}+\omega^2 Q_2^{(S)}}{\omega-\omega^2},\quad
Q_2^{(F)}=\frac{\omega^2Q_1^{(S)}+\omega Q_2^{(S)}}{\omega-\omega^2}.
\label{qfbyqs}
\end{align}
By taking account of $Q_a^{(S)}\in\Gamma_3$ and the constraint (\ref{z3constraint}),
we can show that $Q_a^{(F)}$ are also element of $\Gamma_3$ and
no constraint is imposed.
Furthermore, (\ref{qfbyqs}) is one-to-one.
From the quantization of $Q_a^{(F)}$ we can fix the modulous on the flat background
as $\tau^{(F)}=\omega$.

The other central charge $\ol Z$
can be directly obtained
by the general formula (\ref{zisqz})
only in the flat background,
and in the S-fold we cannot calculate it directly
because it is
generated non-perturbatively.
However, once we obtain $\ol Z$ in the flat background,
we can use the relations (\ref{z1z2byy1y2}) and (\ref{qfbyqs}) to rewrite it
as a function of the variables on the S-fold side.
The result is
\begin{align}
\ol Z=Q_1^{(F)}y_1^*+Q_2^{(F)}y_2^*
=-Q_1^{(S)}z_2^*-Q_2^{(S)}z_1^*.
\label{z3olz}
\end{align}
At the second equality we used (\ref{z1z2byy1y2}) and (\ref{qfbyqs}).
On the S-fold side this is non-local in the sense that
each term contains the charge and the coordinate
of different branes.

For the following discussions the equation
\begin{align}
|Z|^2-|\ol Z|^2=(|Q_1^{(S)}|^2-|Q_2^{(S)}|^2)(|z_1|^2-|z_2|^2)
\end{align}
may be convenient.
By using this
we can easily check that for an electric junction $\bm{w}$
with $Q_2^{(S)}=-Q_1^{(S)*}$ the absolute values of two central charges coincides;
$|\ol Z[\bm{w}]|=|Z[\bm{w}]|$.

As a consistency check let us compare
singularities in the moduli spaces
of the two set-ups.

When $z_1=z_2$, $z_1=\omega z_2$, or
$z_1=\omega^2z_2$,
two D3-branes in the S-fold collide
and the gauge symmetry is enhanced to $U(2)$.
These three singular loci correspond to
$y_1=y_2$, $y_1=0$, and $y_2=0$, respectively,
in the flat background.
On these loci
two of three D3-branes coincide, and the gauge symmetry
is enhanced to $U(2)$, just like on the S-fold side.
This agreement of the loci is rather trivial
because we determine the relations
among moduli parameters by requiring the
coincidence of the central charge $Z$ of
string connecting two independent D3-branes,
which vanishes if two D3-branes collide.

\begin{figure}[htb]
\centering
\includegraphics{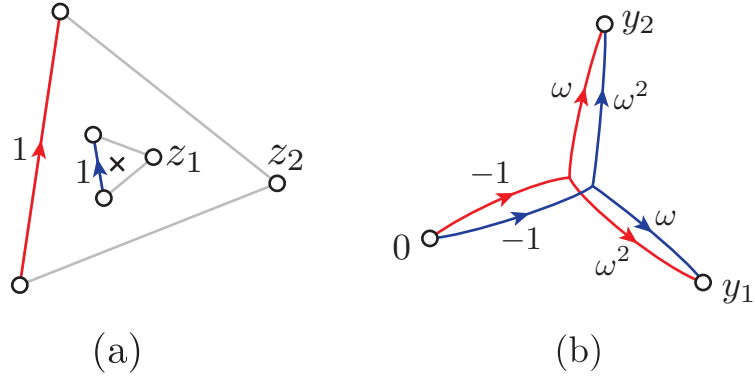}
\caption{Examples of strings in the S-fold and corresponding junctions in the flat background.}
\label{wall.eps}
\end{figure}

On the S-fold side, we have another type of singularity.
When one of $z_1$ and $z_2$ approaches the S-plane,
a string connecting the D3-brane and one of its mirrors
shrinks to zero length,
and appearance of massless particles is naively expected.
This limit, however, causes no singularity
on the flat background.
When $z_1=0$ or $z_2=0$, three D3-branes in the flat background
form an equilateral triangle, and no junction shrinks to zero size.
In Figure \ref{wall.eps} (a) we show two strings
\begin{align}
z_{1,a}\stackrel{1}{\longleftarrow}z_{2,a},\quad(a=1,2),
\end{align}
in a situation with $|z_1|\ll|z_2|$.
The dyonic charges of these strings are
\begin{align}
(Q_1^{(S)},Q_2^{(S)})=\left\{
\begin{array}{ll}
(\omega-\omega^2,0) & (a=1) \\
(0,\omega-\omega^2) & (a=2)
\end{array}.
\right.
\end{align}
By the relation (\ref{qfbyqs}) we obtain the
charges of the corresponding junctions:
\begin{align}
(Q_1^{(F)},Q_2^{(F)})=\left\{
\begin{array}{ll}
(\omega,\omega^2) & (a=1) \\
(\omega^2,\omega) & (a=2)
\end{array}.
\right.
\end{align}
Junctions carrying these charges in the flat background are shown in
Figure \ref{wall.eps} (b).
Although they have similar shapes
the corresponding strings have
completely different length in the S-fold.
In particular, in the $z_1=0$ limit
the junctions in (b) seem to have the same mass
while in (a)
the red one is massive and the blue one
is massless.

One may think that all these strings/junctions are
BPS, and naive estimation of masses should give
the correct values.
If it were the case, the above observation would mean
inequivalence of theories given by two set-ups.
This is, however, too naive.

Remember that for planar junctions on the 89 plane
in the ${\cal N}=4$ theory
there are two central charges $Z$ and $\ol Z$,
which gives independent BPS bounds.
The two junctions in Figure \ref{wall.eps} (b)
are $1/4$ BPS configurations saturating different
bounds;
the red one is a holomorphic junction saturating
$m=|Z|>|\ol Z|$ while the blue one
is an anti-holomorphic junction
saturating $m=|\ol Z|>|Z|$.
On the S-fold, however, only $Z$ is realized perturbatively,
and $\ol Z$ is generated by some unknown non-perturbative
dynamics.
From the length of strings in Figure \ref{wall.eps} (a)
we can determine only $Z$ of two strings.
Although they give the lower bounds for the masses,
the saturation is not guaranteed.
If $|\ol Z|$ is greater than $|Z|$
the bound $m\geq|Z|$ is never saturated,
and this is the case for the blue
string in Figure \ref{wall.eps} (a).
We emphasize that there is no reason that
the perturbative estimation gives good approximation
of the mass.
We consider the situation in which
$z_1$ is much smaller than the string scale
and the string coupling is of order $1$,
and we cannot exclude large quantum corrections.

\subsection{$\ZZ_4$ S-fold}
Next let us consider $\ZZ_4$ case.
The modulus is $\tau=i$ and the complex charge
of strings in the S-fold
takes value in the square lattice $\Gamma_4=\{p+iq|p,q\in\ZZ\}$.
If we put two D3-branes on this background
it is expected to give the
${\cal N}=4$ $SO(5)$ theory,
which can also be realized by an orientifold.
In this subsection we determine the spectrum of junctions
in the S-fold and compare it with the spectrum
on the orientifold side.

Before starting the analysis of the junction spectra
let us briefly review some properties of O3-planes
relevant to the analysis below.
There are four types of O3-planes
distinguished by the discrete torsion
$t\in\ZZ_2+\ZZ_2$
of
R-R and NS-NS three-form fluxes \cite{Witten:1998xy}.
If we put D3-branes in an orientifold ${\cal N}=4$
theory with an orthogonal or
symplectic gauge group is realized.
The relation between the torsion and the gauge group is
shown in Table \ref{o3planes}.
\begin{table}[t]
\caption{The discrete torsions and the gauge groups realized by the orientifolds
with four kinds of O3-planes are shown.
The first and the second component of $t$ represent the
torsions for R-R and NS-NS fluxes, respectively.
$N$ is the number of the mobile D3-branes.}\label{o3planes}
\begin{center}
\begin{tabular}{ccccc}
\hline
\hline
& $\mbox{O3}^-$ & $\wt{\mbox{O3}}^-$
& $\mbox{O3}^+$ & $\wt{\mbox{O3}}^+$ \\
\hline
$t$ & $(0,0)$ & $(1,0)$ & $(0,1)$ & $(1,1)$ \\
$G$ & $SO(2N)$ & $SO(2N+1)$ & $Sp(N)$ & $Sp(N)$ \\
\hline
\end{tabular}
\end{center}
\end{table}
Among four types of O3-planes,
the three but one with $t=(0,0)$ are
transformed among them by $SL(2,\ZZ)$ symmetry
of the type IIB string theory.
Correspondingly, the gauge theories realized by them
are equivalent via Montonen-Olive duality.
In the case with two D3-branes, which we are interested in,
the gauge groups $SO(5)$ and $Sp(2)$ are isomorphic,
and the equivalence of these theories
is trivial.
However, the W-bosons are realized in different ways.
We refer to these two realization of the theory as
$SO(5)$ and $Sp(2)$ theories.

\begin{figure}[t]
\begin{center}
\includegraphics[width=7cm]{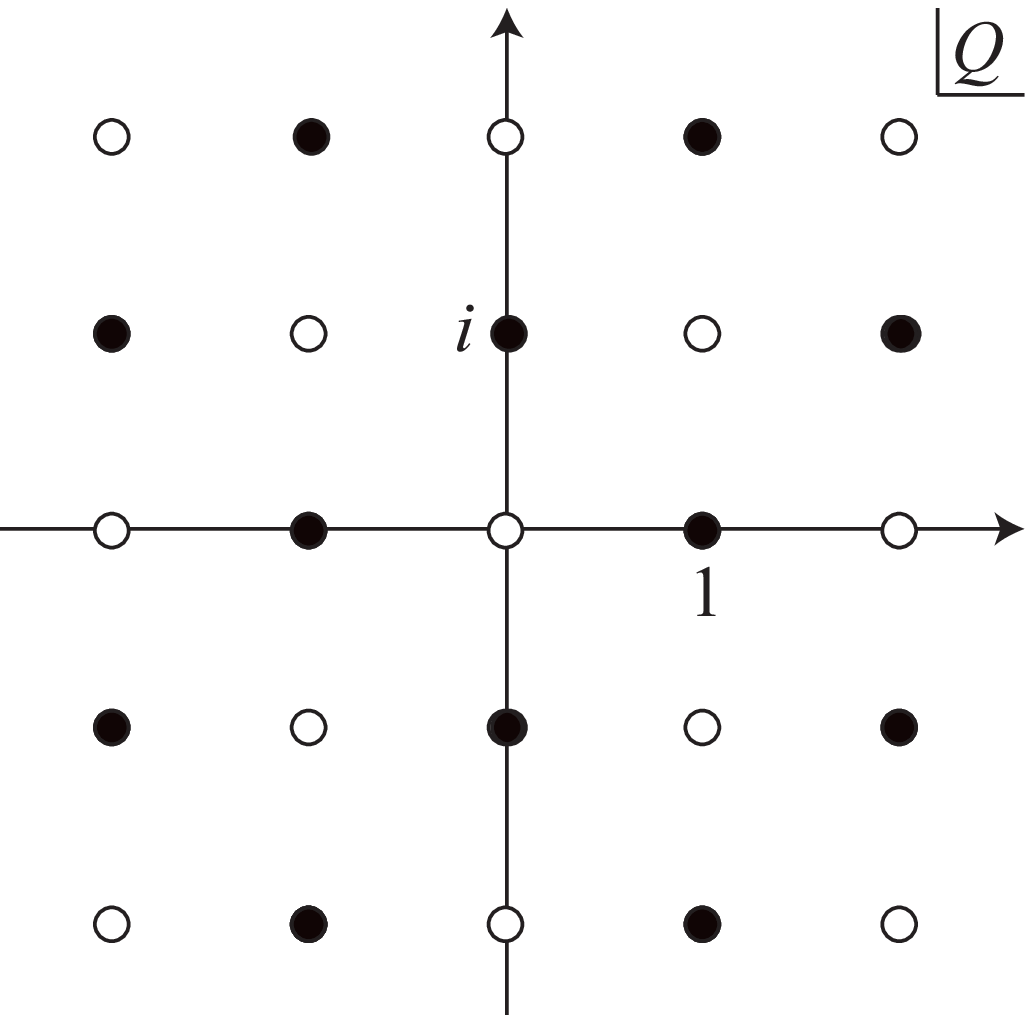}
\caption{A $\ZZ_2$ grading of the lattice $\Gamma_4$ is shown.
White and black dots represent charges with $F(Q)=0$ and $1$, respectively.} \label{ox}
\end{center}
\end{figure}

Let us first determine the junction spectrum in the $\ZZ_4$ S-fold
with two D3-branes.
The charges of strings $Q_a^{(S)}$ ($a=1,2$)
ending on the D3-branes satisfy
\begin{align}
F_4(Q_1^{(S)}+Q_2^{(S)})=0,
\label{z4constraint}
\end{align}
where $F_4$ is the homomorphism from $\Gamma_4$ onto
$\ZZ_2$ which define the $\ZZ_2$ grading of $\Gamma_4$ shown in Figure \ref{ox}.
(\ref{z4constraint}) shows that
the charges $Q^{(S)}_1$ and $Q^{(S)}_2$
must have the same $\ZZ_2$ grading.

The strings whose charges satisfy the electric condition
(\ref{electric}) are
\begin{align}
\bm{a}  : z_1 \stackrel{1}{\longleftarrow} z_2,\quad
\bm{b}  : z_1 \xleftarrow[\hspace{18pt}]{1-i} i z_2,\quad
\bm{c}  : z_1 \stackrel{i}{\longleftarrow} -z_2,\quad
\bm{d}  : z_1 \xleftarrow[\hspace{18pt}]{1+i}-i z_2,
\label{strinz4s}
\end{align}
and ones with opposite orientation.
We would like to identify these strings with W-bosons
of $SO(5)$ and $Sp(2)$ theories
that are realized on the orientifolds.

Let us consider a system
with an O3-plane at $y=0$ and two D3-branes at $y=y_1,y_2$
(and their mirrors at $y=-y_1,-y_2$).
The complex charge of $(p,q)$-string is
$Q^{(O)} = p + q\tau^{(O)}$.
(The superscripts ``$(O)$'' indicate orientifold.)
We should notice that the modulus $\tau^{(O)}$ may not be the
same as $\tau^{(S)}=i$.
This should be determined so that the junction spectrum agrees
with that of the S-fold.
We denote complex charges of strings ending on the D3-branes
by $Q_a^{(O)}$ ($a=1,2$).
Following (\ref{q0}) we also define
\begin{align}
Q^{(O)}_0=Q^{(O)}_1+Q^{(O)}_2\label{Qorient}.
\end{align}
This is the charge of strings ending on the O3-plane.
The general formula (\ref{attachable})
says that
a $(p,q)$ string can end
on an O3-plane with $t\neq0$
only when the charges satisfy
\begin{align}
(p,q)=0\mbox{ or }t\mod2.
\end{align}
In particular, a fundamental string with charge $(1,0)$
can end on $\wt{\mbox{O3}}^-$-plane, but not on O3$^+$- and $\wt{\mbox{O3}}^+$-planes.

The W-bosons of $SO(5)$ and $Sp(2)$ realizations
are given in a consistent way to this constraint.
For an $\wt{\mbox{O3}}^-$-plane
$SO(5)$ W-bosons are given by
open strings
\begin{align}
y_2 \stackrel{1}{\longleftarrow} y_1,\quad
-y_1 \stackrel{1}{\longleftarrow} y_2,\quad
0 \stackrel{1}{\longleftarrow} y_1,\quad
y_2 \stackrel{1}{\longleftarrow} 0,
\label{so5string}
\end{align}
while for O3$^+$- and $\wt{\mbox{O3}}^+$-planes $Sp(2)$ W-bosons
arise as strings
\begin{align}
y_2 \stackrel{1}{\longleftarrow} y_1,\quad
-y_1 \stackrel{1}{\longleftarrow} y_2,\quad
-y_1 \stackrel{1}{\longleftarrow} y_1,\quad
y_2 \stackrel{1}{\longleftarrow} -y_2.
\label{sp2string}
\end{align}
We want to determine relations between
strings (\ref{strinz4s}) in the S-fold
and (\ref{so5string}) or (\ref{sp2string})
in the orientifold.
The first two in (\ref{so5string})
are the same with the first two in (\ref{sp2string}):
\begin{align}
y_2 \stackrel{1}{\longleftarrow} y_1,\quad
-y_1 \stackrel{1}{\longleftarrow} y_2.
\label{common}
\end{align}
Let us start with matching these two with
two of the strings in (\ref{strinz4s}).
Because two strings in (\ref{common}) are continuously deformed to each other
by moving D3-branes, so are the corresponding strings.
Among four strings in (\ref{strinz4s})
$\bm{a}$ and $\bm{c}$
are
deformed to each other,
and so are $\bm{b}$ and $\bm{d}$, too.
Therefore, we have only two essentially different choices.

First let us try matching
strings in (\ref{common}) with $\bm{d}$ and $\bm{b}$.
Matching of these two is sufficient to determine the
relation between the moduli parameters and that between
dyonic charges.
By identifying the central charges $Z^{(S)}$ and $Z^{(O)}$ we obtain
the relations
\begin{align}
z_1 = -\frac{1}{2}(y_1+i y_2),\quad
z_2 = \frac{1}{2}(y_1-i y_2). \label{so5zy}
\end{align}
From the comparison of charges
we obtain
\begin{align}
Q^{(O)}_1=-\frac{1}{2}(Q^{(S)}_1-Q^{(S)}_2),\quad
Q^{(O)}_2=-\frac{i}{2}(Q^{(S)}_1+Q^{(S)}_2). \label{QFQS1}
\end{align}
These relations automatically fix the correspondents of
$\bm{a}$ and $\bm{c}$ in
(\ref{strinz4s}), and we find that
they correspond to the two remaining strings in (\ref{so5string}).
(See Table \ref{chargeso}.)
\begin{table}[htb]
\caption{The dyonic charges and the central charge of junctions in the S-fold and $\wt{\mbox{O3}}^-$-plane background
are shown for four strings corresponding to W-bosons.}
\label{chargeso}
\begin{center}
	\begin{tabular}{ccccc} \hline \hline
	 & $Z^{(S)}$ & $(Q^{(S)}_1,Q^{(S)}_2)$ & $Z^{(O)}$ & $(Q^{(O)}_1,Q^{(O)}_2)$ \\ \hline
	$\bm{a}$ & $z_1-z_2$ & $(1,-1)$ & $-y_1$ & $(-1,0)$ \\
	$\bm{b}$ & $(1-i)(z_1-i z_2)$ & $(1-i,-1-i)$ & $-y_1-y_2$ & $(-1,-1)$ \\
	$\bm{c}$ & $i(z_1+z_2)$ & $(i,i)$ & $y_2$ & $(0,1)$ \\
	$\bm{d}$ & $(1+i)(z_1+i z_2)$ & $(1+i,-1+i)$ & $-(y_1-y_2)$ & $(-1,1)$ \\ \hline
	\end{tabular}
\end{center}
\end{table}
Namely, this gives the correspondence
to the adjoint representation in the $SO(5)$ theory.

\begin{figure}[t]
\begin{center}
\includegraphics[width=7cm,clip]{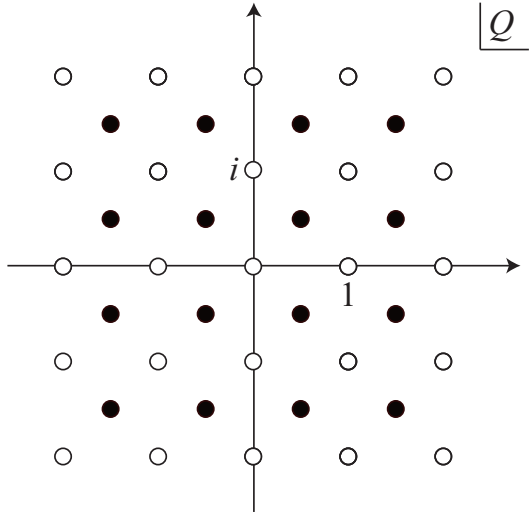}
\caption{A $\ZZ_2$ grading of the lattice with $\tau=(1+i)/2$ is shown.
White and black dots represent charges with $0$ and $1$ in $\ZZ_2$, respectively.} \label{lat}
\end{center}
\end{figure}

$Q_a^{(S)}$ satisfying
(\ref{z4constraint})
can be given by
\begin{align}
Q_a^{(S)}=p_a+iq_a,\quad
p_1+p_2+q_1+q_2\in2\ZZ.
\label{ppqq2z}
\end{align}
Substituting this
into (\ref{QFQS1})
we obtain
\begin{align}
Q^{(O)}_1=\frac{1}{2}[(p_1-p_2)+i(q_1-q_2)],\quad
Q^{(O)}_2=\frac{1}{2}[(-q_1-q_2)+i(p_1+p_2)].
\label{qfforso5}
\end{align}
(\ref{ppqq2z}) implies that these charges take values in
the $\ZZ_2$ graded lattice in Figure \ref{lat},
and $Q^{(O)}_1$ and $Q^{(O)}_2$ have the same grading.
The latter statement means that
$Q_0^{(O)}$, the charge of
the string ending on the O3-plane,
has grading $0$.

From the quantization of each $Q_a^{(O)}$
we can fix the modulous
in the orientifold
to be
\begin{align}
\tau^{(O)} = \frac{1}{2}(1+i).
\end{align}
Then $Q_a^{(O)}$ are given in the form $p+q\tau$,
and $Q_0^{(O)}$, which has trivial grading,
can be expressed as
\begin{align}
Q_0^{(O)}=p+2q\tau^{(O)},\quad p,q\in\ZZ.
\end{align}
This is nothing but the condition (\ref{attachable})
for an $\wt{\mbox{O3}}^-$-plane, which gives
the gauge group $SO(5)$.

Let us consider the other possibility:
matching of two strings in (\ref{common})
with $\bm{a}$ and $\bm{c}$
in (\ref{strinz4s}).
We obtain the relations
\begin{align}
z_1 = -\frac{1}{2}(1-i)y_1+\frac{1}{2}(1+i)y_2, \hspace{7pt} 
z_2 = \frac{1}{2}(1+i)y_1-\frac{1}{2}(1-i)y_2,\label{sp2zy}
\end{align}
and
\begin{align}
Q^{(O)}_1=-\frac{1-i}{2}Q^{(S)}_1 + \frac{1+i}{2}Q^{(S)}_2,\quad
Q^{(O)}_2=\frac{1+i}{2}Q^{(S)}_1 - \frac{1-i}{2}Q^{(S)}_2 \label{QFQS2},
\end{align}
and the correspondents of $\bm{b}$ and $\bm{d}$ are
the two remaining strings in (\ref{sp2string}).
(See Table \ref{chargesp}.)
\begin{table}[htb]
\caption{The dyonic charges and the central charge of junctions in the S-fold and 
O3$^+$- and $\wt{\mbox{O3}}^+$-planes background
are shown for four strings corresponding to W-bosons.}
\label{chargesp}
\begin{center}
	\begin{tabular}{ccccc} \hline \hline
	 & $Z^{(S)}$ & $(Q^{(S)}_1,Q^{(S)}_2)$ & $Z^{(O)}$ & $(Q^{(O)}_1,Q^{(O)}_2)$ \\ \hline
	$\bm{a}$ & $z_1-z_2$ & $(1,-1)$ & $-(y_1-y_2)$ & $(-1,1)$ \\
	$\bm{b}$ & $(1-i)(z_1-i z_2)$ & $(1-i,-1-i)$ & $2y_2$ & $(0,2)$ \\
	$\bm{c}$ & $i(z_1+z_2)$ & $(i,i)$ & $-y_1-y_2$ & $(-1,-1)$ \\
	$\bm{d}$ & $(1+i)(z_1+i z_2)$ & $(1+i,-1+i)$ & $-2y_1$ & $(-2,0)$ \\ \hline
	\end{tabular}
\end{center}
\end{table}

Substituting (\ref{ppqq2z}) into 
(\ref{QFQS2}) we obtain
\begin{align}
Q_1^{(O)}=\frac{1+i}{2}[(p_2-q_1)+i(p_1+q_2)],\quad
Q_2^{(O)}=\frac{1+i}{2}[(p_1-q_2)+i(p_2+q_1)].
\end{align}
These are similar to (\ref{qfforso5})
and only difference
up to unimportant signatures
are the extra factors $1+i$.
Therefore,
the quantization and constraint for
$Q_a^{(O)}$ can be obtained
from the previous ones
by simply rotate
the lattice in Figure \ref{lat} by $45$ degrees
and expand it by the factor $\sqrt2$.
The resulting lattice is the same as
Figure \ref{ox}.
If we set $\tau^{(O)}=1+i$ the charges
$Q_a^{(O)}$ take the form $p+q\tau^{(O)}$,
and $Q_0^{(O)}$ can be expressed as
\begin{align}
Q_0^{(O)}=2p+q\tau^{(O)},\quad p,q\in\ZZ.
\end{align}
This is the constraint
(\ref{attachable}) for O3$^+$-plane.
We can also set $\tau^{(O)}=i$.
Then $Q^{(O)}_0$ can be expressed as
\begin{align}
Q_0^{(O)}=2p+q(1+\tau^{(O)}),\quad
p,q\in\ZZ.
\end{align}
This is the constraint
(\ref{attachable}) for $\wt{\mbox{O3}}^+$-plane.
These two assignments of $\tau^{(O)}$ give
two $Sp(2)$ theories with different $\theta$-angle,
which are dual to each other.

In both cases, the spectrum of the central charge $Z$
in the S-fold
matches completely with that of the orientifold.

Once we obtain these relations, we can rewrite $\ol Z$ as a function of the S-fold variables.
For the $SO(5)$ matching with (\ref{so5zy}) and (\ref{QFQS1}) and
the $Sp(2)$ matching with (\ref{sp2zy}) and (\ref{QFQS2}) we obtain
\begin{align}
SO(5):\ol Z= Q_1^{(S)}z_2^*+Q_2^{(S)}z_1^*,\quad
Sp(2):\ol Z= -Q_1^{(S)}z_2^*-Q_2^{(S)}z_1^*.
\end{align}
The overall sign depends on the choice of string orientation
and is the matter of conventions.
This result is the same as (\ref{z3olz}) up to sign.

\begin{figure}[htb]
\centering
\includegraphics{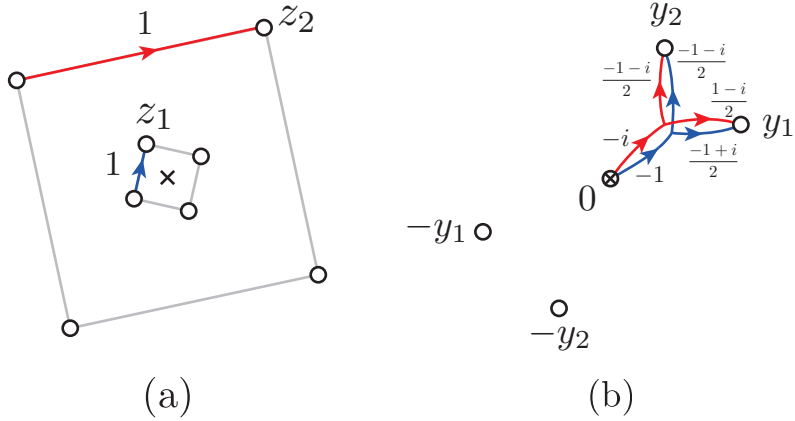}
\caption{Examples of strings in the S-fold and corresponding junctions in the orientifold background.}
\label{z4wall.eps}
\end{figure}

As in the $k=3$ case
let us illustrate the junctions corresponding to strings in S-fold.
Strings connecting two independent D3-branes give W-bosons and they have been already analyzed
above, and here we focus on strings connecting a D3-brane
and one of its mirror branes.
First we consider two strings
\begin{align}
 z_{0,a}\stackrel{1}{\longleftarrow}z_{1,a},\quad(a=1,2).
 \label{eq:1}
\end{align}
These are shown in
Figure \ref{z4wall.eps} (a)
in a situation with $|z_1|\ll |z_2|$.
The dyonic charges of these strings are
\begin{align}
(Q_1^{(S)},Q_2^{(S)})=\left\{
\begin{array}{ll}
(1-i,0) & (a=1) \\
(0,1-i) & (a=2)
\label{z4juncq1q2}
\end{array}.
\right.
\end{align}
By the relation (\ref{QFQS1}) we obtain the
charges of the corresponding junctions:
\begin{align}
(Q_1^{(O)},Q_2^{(O)})=\left\{
\begin{array}{ll}
(\frac{-1+i}{2},\frac{-1-i}{2}) & (a=1) \\
(\frac{1-i}{2},\frac{-1-i}{2}) & (a=2)
\end{array}.
\right.
\end{align}
Junctions carrying these charges are shown in
Figure \ref{z4wall.eps} (b).
As in Figure \ref{wall.eps}
the red and blue junctions in Figure
\ref{z4wall.eps} (b) are
holomorphic
and
anti-holomorphic, respectively.

Another pair of two strings in S-fold is
\begin{align}
z_{0,a}\stackrel{1}{\longleftarrow}z_{2,a},\quad(a=1,2).
\end{align}
We can easily check that these strings
carry dyonic charges
that are as $1+i$ times as those in (\ref{z4juncq1q2}).
The corresponding junctions
are ones in Figure \ref{z4wall.eps} (b) with
charges multiplied by $1+i$.

\subsection{$\ZZ_6$ S-fold}
For $k=6$ we expect ${\cal N}=4$ SYM with the
gauge group $G_2$.
The discrete torsion group for $k=6$ is trivial,
and an arbitrary pair of the dyonic charges
$(Q^{(S)}_1,Q^{(S)}_2)\in\Gamma_6+\Gamma_6$
is allowed.
The central charge $Z$ is given by
\begin{align}
Z=Q^{(S)}_1z_1+Q^{(S)}_2z_2.
\label{zqzz}
\end{align}
Although no perturbative realization
of $G_2$ theory is known in string theory
and we cannot directly read off the
other central charge $\ol Z$
from junctions,
it is natural
to guess
from the results for $k\leq 4$
that
\begin{align}
\ol Z=Q^{(S)}_1z_2^*+Q_2^{(S)}z_1^*,
\label{olzqz}
\end{align}
up to overall phase.
Indeed, if we assume
\begin{itemize}
\item
$Z$ is given by (\ref{zqzz}),
\item
$\ol Z$ is a holomorphic bilinear form of $Q^{(S)}_a$ and $z_a^*$.
Namely, $\ol Z$ is given by $\ol Z=c_{ab}Q_a^{(S)}z_b^*$
with some coefficients $c_{ab}$, and
\item
$|Z|=|\ol Z|$ for charges satisfying the electric condition
(\ref{electric}),
\end{itemize}
then (\ref{olzqz}) is the unique solution
up to phase ambiguity.

\section{Walls of marginal stability}\label{wall.sec}
In the previous sections we show that there are two types of
$1/4$ BPS junctions:
holomorphic and anti-holomorphic ones.
They may decay or make the transition between two types of BPS states as we move D3-branes.
In this section we demonstrate how this
happens for some junctions.
The purpose of this section is not to give a comprehensive analysis of
general junctions but simply to show typical processes that
occur for simple junctions when
we move D3-branes. Thus, we consider only strings in S-folds
that are shown in Figure~\ref{wall.eps} and Figure~\ref{z4wall.eps}.

\begin{figure}[htb]
\centering
\includegraphics{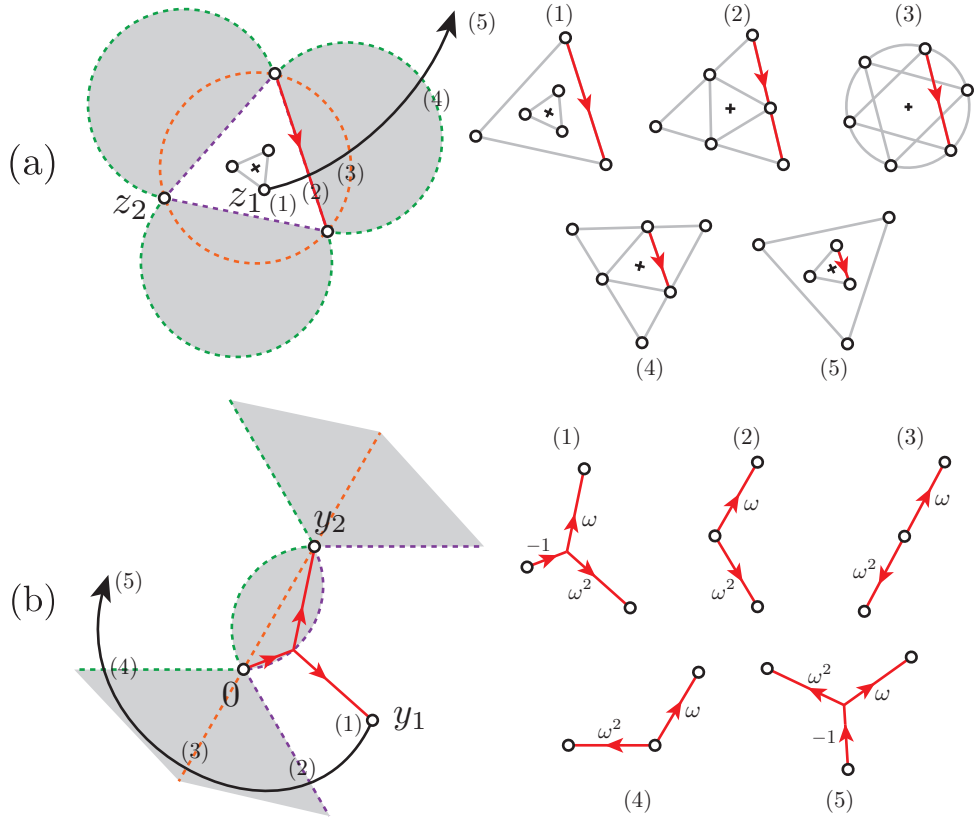}
 \caption{
 Moduli spaces for $\ZZ_3$ case are shown.
 $z_2$, $y_2$, and the origin are fixed and $z_1$ and $y_1$ moves.
 Walls, regions, and a typical example of wall crossing are illustrated.
 }
\label{moduli.eps}
\end{figure}

Let us consider $\ZZ_3$ case first.
As a starting point
we take the red string and junction in Figure~\ref{wall.eps},
which are also shown in Figure \ref{moduli.eps}.
As we previously mentioned the junction is holomorphic.
Let $y_*$ be the intersection point of the junction.
If $y_*$ coincide $0$, $y_1$, or $y_2$, the junction
decays into a pair of strings.
Let us first consider the wall with $y_*=0$.
Because the angles at the junction point are all $2\pi/3$,
$y_* = 0$ is
equivalent to
\begin{align}
y_2=t\omega y_1,\quad
t\in \RR_+  .
\end{align}
Using (\ref{z1z2byy1y2})
we can rewrite this as the following relation between $z_1$ and $z_2$:
\begin{align}
z_1=\frac{t}{t+1}\omega z_2+\frac{1}{t+1}\omega^2 z_2  ,
\end{align}
which means $z_1$ is an internally dividing point of $\omega z_2$ and $\omega^2 z_2$.
If we illustrate relative position of D3-branes in the S-fold
so that the position $z_2$ is fixed,
this gives a wall of the transition
on the $z_1$-plane.
The other two walls $y_*=y_1$ and $y_*=y_2$
are given respectively by
\begin{align}
-y_1=t\omega(y_2-y_1)
\quad&\leftrightarrow\quad
z_1=\frac{t}{t+1}z_2+\frac{1}{t+1}\omega z_2 ,  \\
y_1-y_2=t\omega(-y_2)
\quad&\leftrightarrow\quad
z_1=\frac{t}{t+1}\omega^2z_2+\frac{1}{t+1}z_2 .
\end{align}
These three walls form
the triangle with vertices at $z_2$, $\omega z_2$, and $\omega^2 z_2$,
which is shown as purple dashed lines in Figure~\ref{moduli.eps} (a).
The corresponding walls in the flat background are shown in (b) by the same color of dashed lines.
When $y_1$($z_1$) crosses these walls
the junction decays into two strings (see (b-2) in Figure~\ref{moduli.eps}).

Let us continue the deformation.
When
$y_2 = a y_1  \quad (a \in \RR)$
all D3-branes in the flat background are aligned on a line,
and then the relation
$|z_1| = |z_2|$ holds, which means
D3-branes in the S-fold are on a circle.
(See orange dashed lines and $(3)$ in Figure~\ref{moduli.eps}.)
Nothing special happens on this locus.

As $z_1$ goes further away from the origin,
finally, it crosses
the walls
defined by
\begin{align}
 y_1=t\omega y_2
\quad&\leftrightarrow\quad
z_1 = \omega^2 z_2 \frac{t+1}{t+\omega} ,  \\
y_2-y_1=t\omega(-y_1)
\quad&\leftrightarrow\quad
z_1 = \omega z_2 \frac{t+1}{t+\omega} ,  \\
-y_2=t\omega(y_1-y_2)
\quad&\leftrightarrow\quad
z_1 = z_2 \frac{t+1}{t+\omega} ,
\end{align}
where $t$ is, again, real positive.
These walls are shown 
as green dashed lines in Figure~\ref{moduli.eps}.
Note that these equations also represent that $z_2$ is on a side of
the triangle $z_1$, $\omega z_1$, and $\omega^2 z_1$.
After the crossing the pair of strings form an
anti-holomorphic junction.
(See (b-5) in Figure~\ref{moduli.eps}.)

Having illustrated the walls,
it is clear that
the gray regions in Figure~\ref{moduli.eps} are
regions where the junctions cannot be sustained,
hence, they are regions of non-BPS states.
The other regions
are ones of holomorphic or anti-holomorphic BPS states.

\begin{figure}[htb]
\centering
\includegraphics{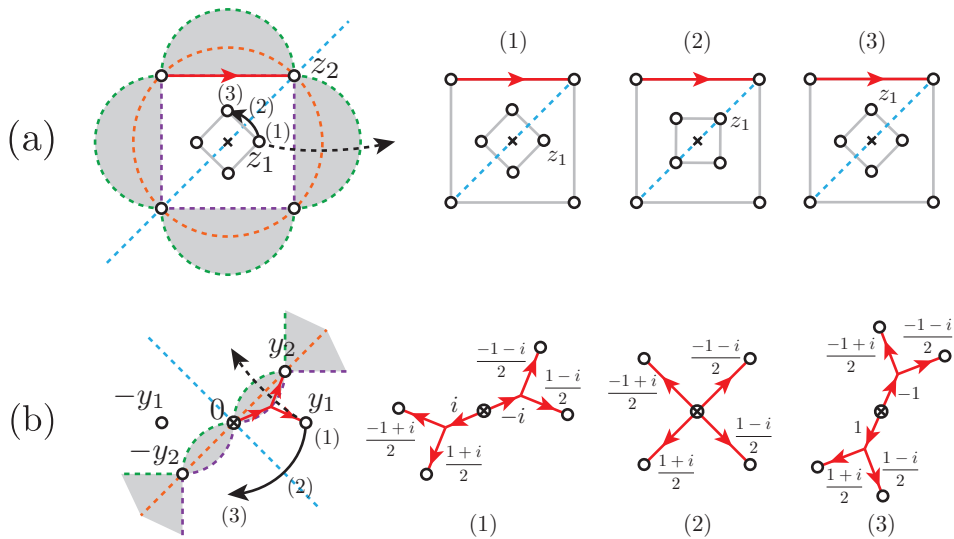}
 \caption{
 Moduli space for a string in $\ZZ_4$ S-fold is shown.
 $z_2$, $y_2$, and the origin are fixed and $z_1$ and $y_1$ moves.
 In the shaded regions the junction becomes non-BPS.
 The D3-brane motion shown by dashed arrows cause a deformation
of the junction similar to what shown in Figure \ref{moduli.eps}.
Some string configurations in the motion along solid arrows
are shown on the right.
 }
\label{z4moduli.eps}
\end{figure}

Let us move on to $\ZZ_4$ case as another example.
We look into the red string and junction illustrated in Figure~\ref{z4wall.eps}.
There are two different points compared to $\ZZ_3$ case.
One is that the flat background in $\ZZ_3$ case is replaced by the orientifold.
The other is that the angles at the junction point are not all $2\pi/3$ but $\pi/2$, $3\pi/4$, and $3\pi/4$.
By taking these into account
we can easily determine the walls,
and the results are illustrated in Figure~\ref{z4moduli.eps}%
\footnote{The analysis here is classical, and we simply assume the existence
of states saturating the BPS bound for $3$-pronged junctions in the orientifold.}.

Let us follow a typical motion of D3-branes.
We move the D3-brane at $z_1$ with $z_2$ kept intact.
If we move $z_1$ outward as shown in Figure~\ref{z4moduli.eps} (a) by the dashed arrow,
the deformation of the corresponding junction is quite similar to that of $\ZZ_3$ case,
and we do not repeat the analysis for it.
We comment on another type of wall
that we do not meet in the $\ZZ_3$ case.
It is illustrated as light-blue lines
in Figure~\ref{z4moduli.eps}.
Let us move D3-branes following
the solid arrows in the figure.
In (b) we show both the junction and its mirror image
by the reason which will become clear shortly.
On the light-blue line each of the junction and its mirror becomes a pair of strings as in (b-2).
If the brane pass through the light-blue line two strings ending on $y_1$ and $-y_2$
form a new junction.
Although the topology of the junction changes in this process
the junction keeps itself holomorphic.

\section{Discussions}\label{discussions.sec}
In this paper we investigated
junctions in S-folds for the purpose of
understanding the supersymmetry enhancement
proposed by Aharony and Tachikawa.
We determined the spectrum of dyonic charges
of the theory from the information of junctions
in the S-fold,
and in the $k=3$ and $k=4$ cases
we confirmed that the spectra are the same as
those of ${\cal N}=4$ $SU(3)$ and $SO(5)$
gauge theories by using another perturbative
brane realization of each theory.
This agreement seems highly non-trivial,
and gives a strong support to the
realization of ${\cal N}=4$ supersymmetry.

There are many unsolved problems.
What is the most desired
would be a direct determination of
the non-perturbative central charges $\ol Z$.
Due to the lack of
the information of $\ol Z$
we could not directly determine the masses
of BPS states.
When we established the correspondence of the
spectrum of junctions and ${\cal N}=4$
dyonic particles, we use the information of
one of the central charge $Z$,
which can be seen perturbatively
in the S-fold.
However, to determine BPS saturating masses,
we also need the other central charge $\ol Z$.
Because this central charge is generated
non-perturbatively together with
the fourth supercharge in the supersymmetry enhancement,
we cannot determine it by
simply drawing the shapes of junctions.

There is another
problem related to the marginal deformation.
In this paper we showed that the spectrum of
S-fold side gives dyonic spectrum of ${\cal N}=4$ theory
with a particular value of the marginal deformation parameter
$\tau$.
However, if ${\cal N}=4$ is realized, we
should be able to freely change
the deformation parameter.
The change of the parameter
affects the central charges,
and we should have the corresponding parameter
on the S-fold side.
The identification of the
parameter in the S-fold is
very important problem to understand
the supersymmetry enhancement.

Even if we could obtain the central charges,
it would be another problem
to determine the BPS spectrum.
The existence of BPS saturating states
is highly non-trivial problem
and we need to perform quantum analysis
of the junctions to determine it.
As far as we know this is open problem
even for orientifolds.

In Section \ref{wall.sec} we studied wall crossing
for simple string configurations.
We determined walls by using another brane realization
of the theory which is expected to be equivalent to the S-fold system.
It is still unknown what happens on the S-fold
side when the wall is crossed.

We believe any of these questions is
important to understand dynamics of
${\cal N}=3$ theories and S-folds.
We hope to return to these problems in future.

\section*{Acknowledgements}
We would like to thank Tetsuji Kimura for valuable discussions.
The work of YI is partially supported by Grand-in-Aid for Scientific Research (C) (No.15K05044),
Ministry of Education, Science and Culture, Japan.
DY is supported by the ERC Starting Grant N. 304806, ``The Gauge/Gravity Duality and Geometry in String Theory''.

\end{document}